\documentclass{article}
\usepackage{spconf,amsmath,graphicx}
\usepackage{booktabs} 
\usepackage{cite} 
\usepackage{amssymb} 
\usepackage{stfloats} 
\usepackage{hyperref} 

\usepackage[table, dvipsnames]{xcolor}

\definecolor{vlightgray}{gray}{0.95}

\title{Overview of the 2023 ICASSP SP Clarity Challenge: Speech Enhancement for Hearing Aids}

\name{\begin{tabular}{c}Trevor J. Cox$^1$, Jon Barker$^2$, Will Bailey$^{1,2}$, Simone Graetzer$^1$, \\Michael A. Akeroyd$^3$, John F. Culling$^4$, Graham Naylor$^3$\thanks{This work is funded by the UK's Engineering and Physical Sciences Council (EP/S031448/1, S031308/1, S031324/1 and S030298/1). We are grateful to Amazon, the Hearing Industry Research Consortium and the Royal National Institute for the Deaf (RNID) for their
support. 
}\end{tabular}}

\address{
  $^1$ Acoustics Research Centre, University of Salford, UK\\
  $^2$ Department of Computer Science, University of Sheffield, UK\\
  $^3$ School of Medicine, University of Nottingham, UK\\
  $^4$ School of Psychology, Cardiff University, UK
}

\begin{document}
\ninept

\maketitle
%


\begin{abstract}
This paper reports on the design and outcomes of the ICASSP SP Clarity Challenge: Speech Enhancement for Hearing Aids. The scenario was a listener attending to a target speaker in a noisy, domestic environment. There were multiple interferers and head rotation by the listener. The challenge extended the second Clarity Enhancement Challenge (CEC2) by fixing the amplification stage of the hearing aid; evaluating with a combined metric for speech intelligibility and quality; and providing two evaluation sets, one based on simulation and the other on real-room measurements. Five teams improved on the baseline system for the simulated evaluation set, but the performance on the measured evaluation set was much poorer. Investigations are on-going to determine the exact cause of the mismatch between the simulated and measured data sets. The presence of transducer noise in the measurements, lower order Ambisonics harming the ability for systems to exploit binaural cues and the differences between real and simulated room impulse responses are suggested causes.
\end{abstract}

\section{Introduction}



\begin{figure*}[htb]
\centering
\centerline{\includegraphics[width=178 mm]{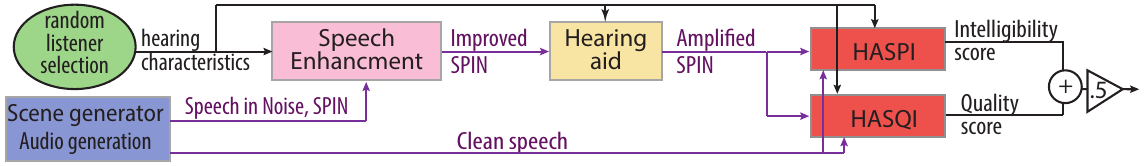}}
\caption{\label{fig:baseline} The baseline system provided.}
\end{figure*}

The WHO estimates that 430 million people have a disabling impairment at a global cost of \$750 billion. Hearing aids are the main treatment for hearing loss, but these these do not work well for some speech-in-noise scenarios. Recent advances in multichannel speech enhancement using deep-learning \cite{fei2020} offer the possibility of better hearing aid processing. Research is also now providing the low-power and low-latency approaches needed for hearing aids (e.g., \cite{l3das22}).

In 2021, a new series of open machine-learning challenges for speech intelligibility enhancement for hearing aids was launched. The ICASSP 2023 SP Clarity Challenge (ClarityICASSP) \footnote{Clarity Challenge, \url{https://claritychallenge.org}} was the third in the series. This built upon the 2\textsuperscript{nd} Clarity Enhancement Challenge (CEC2) \cite{CEC2}.

Both CEC2 and ClarityICASSP provided teams with hearing aid inputs, and asked entrants to enhance the signals for listeners with hearing loss quantified by audiograms. The scenes were typical living rooms with up to three competing interferers of music, speech and domestic noise. Head rotation by the listener simulated a person turning towards the talker. Target talker enrollment sentences were provided, so that systems could learn which speech to attend to.

For CEC2, entries were first evaluated using the objective intelligibility metric HASPI (Hearing Aid Speech Perception Index)\cite{kates2014}. For ClarityICASSP, the entrants were evaluated using the average of two objective metrics: HASPI and HASQI (Hearing Aid Speech Quality Index)\cite{hasqi}. HASQI was added as some entrants to CEC2 had improved intelligibility at the expense of naturalness.

In CEC2, data was created using the following simulation. Dry recordings of speech and noise were convolved with binaural room impulse responses from a geometric room acoustic model and applying HRTFs. A key research aim for ClarityICASSP was to test how well systems generalised beyond simulation. To do this a second evaluation test set was created using more ecologically-valid recordings. A final novelty for ClarityICASSP was to fix the hearing aid amplification stage (yellow box in Fig. \ref{fig:baseline}) \cite{byrne1986national}. This meant researchers less familiar with hearing loss could concentrate on the speech enhancement (pink box).


A companion paper on CEC2 \cite{CEC2} details the scenarios: target utterance, interferers, head rotation, room geometry and layout. Below, we focus on the construction of the new measured evaluation set and the results of the ClarityICASSP challenge.

\section{More ecologically-valid evaluation set}
\label{s:eval2}

The Eval2 data set used recordings from live actors in a listening room (mid-freq. rev. time 0.27 s; 6.6×5.8×2.8 m; 5.7 dBA background noise). The positions of the sources and listeners were chosen using the same methods as for the simulated data.

Recording: (i) The actors were recorded on a Neumann KM184 cardioid mic at 50 cm and a 1\textsuperscript{st}-order Sennheiser AMBEO VR Ambisonic mic at the listener position. This was done in noise-free conditions. (ii) Noise, music and speech interferers were later played from a M-audio BX8a loudspeaker and recorded on the Ambisonic mic.

Post-processing: (i) Head rotations were done using the spherical harmonic representation of the sound. (ii) HRTFs were used to get the hearing-aid microphone signals. (iii) The target talker and interferers were mixed to the desired signal-to-noise ratio.

The 1,600 new sentences were selected from the British National Corpus using the same process as before \cite{ClaritySentences}. These were read by 5 male and 5 female actors whose ages ranged from 20 to 62. They were standing, and told to face and talk to the Ambisonic mic. Each recorded 160 unique sentences, in 10 talking positions. The close cardioid mic was then the reference speech for HASPI and HASQI.

\section{Submissions and Results}
\label{s:submissions}
    
9 systems were submitted by 7 teams. The results are summarised in Table ~\ref{table:results}. For the simulated Eval 1 data, five teams had entries that improved on the baseline, with two producing worse scores. The HASPI and HASQI values were highly correlated (correl. coeff. r = 0.943, using best entry from each team). Across the successful systems, the improvement in the HASQI quality scores were about half the improvement in HASPI intelligibility scores.

The more ecologically valid Eval 2 set produced lower scores for both objective measures across all teams. While four teams still managed to beat the baseline, the improvement was much smaller than for the simulated Eval 1 data.

\section{Discussions and Conclusions}
\label{s:conclusions}

The challenge demonstrated that modern speech enhancement methods can provide better signals for a simple hearing aid to amplify. However, the improvement is marginal for the more ecologically-valid evaluation set based on real-room recording. A mismatch between the simulated and measured data has harmed the speech enhancement processing that used machine learning.

The differences between simulated and ecologically-valid data sets were: (i) Talkers will speak differently when asked to talk to a distant microphone (for the simulation data, they were close miked in a studio). (ii) The room impulse responses of the listening room are naturally different to the approximate simulation from the geometric model. (iii) The directivity of interferers was omnidirectional in the simulation, but had the directivity of the loudspeaker in the listening room recordings. (iv) There was electronic transducer noise on the distant Ambisonic microphone. (v) The measurements used first-order Ambisonics, whereas the simulations used sixth-order.

The exact cause of the mismatch is still being investigated. Of the differences listed above, three seem most likely to be problematic. (i) Transducer noise, as the on-set timing of this was different to interfering noises in the training set. (ii) The first order Ambisonic recordings, as this would have made it harder for the speech enhancement to exploit binaural cues for noise suppression. (iii) Differences between real and simulated room impulse responses.

Finally, while the measured evaluation data was more ecologically-valid, it was not entirely like a real-life situation as the speech and noise were recorded separately, and Ambisonic recording was used with virtual head rotation. Creating evaluation sets using measurements on real hearing aid microphones is therefore planned for Clarity's next enhancement challenge.

\begin{table}
    \centering
    \small
    \caption{ Results from submitted systems. Eval 1 is the simulated evaluation data. Eval 2 is the measured evaluation data. Ave is the mean of the HASPI and HASQI scores. E028d used additional data and E029r used head rotation information. E01 - baseline}\label{table:results}
    \medskip
    \begin{tabular}{lcccccc}
    \toprule
    \multicolumn{1}{c}{ } & \multicolumn{3}{c}{Eval1} & \multicolumn{3}{c}{Eval2} \\
    \hline
    Entry & Ave & HASPI & HASQI & Ave & HASPI & HASQI \\
    \midrule
    E02 & 0.136 & 0.179 & 0.093 & 0.09 & 0.101 & 0.078 \\
\rowcolor{vlightgray} E09 & 0.224 & 0.286 & 0.161 & 0.117 & 0.126 & 0.108 \\
E14 & 0.606 & 0.797 & 0.414 & 0.201 & 0.291 & 0.110 \\
\rowcolor{vlightgray} E23 & 0.082 & 0.117 & 0.047 & 0.018 & 0.026 & 0.009 \\
E28 & 0.653 & 0.78 & 0.526 & 0.022 & 0.026 & 0.019 \\
\rowcolor{vlightgray} E28d & 0.693 & 0.816 & 0.57 & 0.199 & 0.249 & 0.154 \\
E29 & 0.613 & 0.835 & 0.393 & 0.18 & 0.256 & 0.104 \\
\rowcolor{vlightgray} E29r& 0.616 & 0.838 & 0.393 & 0.18 & 0.256 & 0.103 \\
E30 & 0.522 & 0.729 & 0.316 & 0.208 & 0.284 & 0.132 \\
\rowcolor{vlightgray} E01 & 0.197 & 0.266 & 0.128 & 0.149 & 0.176 & 0.121 \\
    \bottomrule    
    \end{tabular}

    \vspace*{-5mm}
    \end{table}

\bibliographystyle{IEEEbib}
\bibliography{strings,refs}

\end{document}